\documentclass[twocolumn,a4paper,superscriptaddress,reprint,aps,prl]{revtex4-1}

\usepackage{amsmath}
\usepackage{graphicx}
\usepackage[english]{babel}

\begin{document}
\title{Excitons and stacking order in h-BN }

\author{Romain Bourrellier}
\affiliation{Laboratoire de Physique des Solides, Univ. Paris-Sud, CNRS UMR
8502, F-91405, Orsay, France}

\author{Michele Amato}
\affiliation{Laboratoire de Physique des Solides, Univ. Paris-Sud, CNRS UMR
8502, F-91405, Orsay, France}
\affiliation{Laboratoire des Solides Irradi\'es, Ecole Polytechnique, Route de
Saclay, F-91128 Palaiseau and European Theoretical Spectroscopy Facility 
(ETSF), France}

\author{Luiz Henrique Galv\~ao Tizei}
\affiliation{Laboratoire de Physique des Solides, Univ. Paris-Sud, CNRS UMR
8502, F-91405, Orsay, France}

\author{Christine Giorgetti}
\affiliation{Laboratoire des Solides Irradi\'es, Ecole Polytechnique, Route de
Saclay, F-91128 Palaiseau and European Theoretical Spectroscopy Facility 
(ETSF), France}

\author{Alexandre Gloter}
\affiliation{Laboratoire de Physique des Solides, Univ. Paris-Sud, CNRS UMR
8502, F-91405, Orsay, France}

\author{Malcolm I. Heggie}
\affiliation{Department of Chemistry, University of Surrey, Guildford GU2 
7XH, United Kingdom}

\author{Katia March}
\affiliation{Laboratoire de Physique des Solides, Univ. Paris-Sud, CNRS UMR
8502, F-91405, Orsay, France}

\author{Odile St\'ephan}
\affiliation{Laboratoire de Physique des Solides, Univ. Paris-Sud, CNRS UMR
8502, F-91405, Orsay, France}

\author{Lucia Reining}
\affiliation{Laboratoire des Solides Irradi\'es, Ecole Polytechnique, Route de
Saclay, F-91128 Palaiseau and European Theoretical Spectroscopy 
Facility (ETSF), France}

\author{Mathieu Kociak}
\affiliation{Laboratoire de Physique des Solides, Univ. Paris-Sud, CNRS UMR
8502, F-91405, Orsay, France}

\author{Alberto Zobelli}
\email{alberto.zobelli@u-psud.fr}
\affiliation{Laboratoire de Physique des Solides, Univ. Paris-Sud, CNRS UMR 
8502, F-91405, Orsay, France}

\begin{abstract}
The strong excitonic emission at 5.75 eV of hexagonal boron nitride (h-BN) makes 
this material one of the most promising candidate for light emitting devices in 
the far ultraviolet (UV). However, single excitons occur only in perfect monocrystals that are 
extremely hard to synthesize, while regular h-BN samples present a complex emission spectrum with 
several additional peaks.
The microscopic origin of these additional emissions has not yet been understood. 
In this work we address this problem using an experimental and theoretical 
approach that combines  nanometric resolved cathodoluminescence, high resolution 
transmission electron microscopy and state of the art theoretical spectroscopy 
methods. We demonstrate that emission spectra are strongly inhomogeneus within 
individual flakes and that additional excitons occur at structural deformations, 
such as faceted plane folds, that lead to local changes of the h-BN stacking order. 
\end{abstract}

\maketitle

In the last years, hexagonal boron nitride (h-BN) emerged as a promising alternative candidate for 
optoelectronic applications in the far UV region \cite{Watanabe2004,Wirtz2005}. An 
efficient BN based ultraviolet (UV) light emitter in the 5.3-5.9 eV energy range has been 
recently obtained using accelerated electrons as the pumping 
source \cite{Watanabe2009b}. For the design of future h-BN devices it is 
fundamental to have an in-depth description and understanding of the complex 
luminescence of this material.

Theoretical and experimental works have demonstrated that the emission spectrum 
of h-BN is dominated by a Frenkel type exciton at 5.75 eV 
\cite{Watanabe2004,Wirtz2005}. However, a single peak emission appears only in 
high quality macroscopic crystals that can be obtained only in a limited amount 
through high pressure and high temperature synthesis processes 
\cite{Kubota2007}. Common h-BN samples present a more complex emission spectrum 
with a series of sharp emissions close to the main exciton in the 5.3-5.9 eV 
energy range. An additional broad emission occurs within the electronic band gap 
in the energy range 3.2-4.5 eV on which a series of three sharp lines might be 
superposed depending on the BN sample purity \cite{Museur2007,Museur2008}.

The origin of this complex emission pattern is still controversial and it has 
been attributed to the presence of unidentified structural defects. Impurities 
have been proposed to be responsible for  emission lines in the band 
gap \cite{Museur2008} and generic dislocations or grain boundaries to be at the 
origin of emission lines near the free exciton. The latter hypothesis has been 
suggested by the appearance of additional emission peaks  when a pristine 
perfect crystal was deformed by a limited mechanical strain 
\cite{Watanabe2006,Watanabe2006a}. Furthermore, it has been shown through 
sub-micrometric emission maps obtained by cathodoluminescence (CL) filtered 
images that the standard exciton is homogeneous in a h-BN crystallite whereas 
additional peaks present some spatial localization 
\cite{Jaffrennou2007a,Jaffrennou2008a}. However, research conducted  in the last 
ten years has not yet been able to correlate the optical properties of h-BN with 
the microscopic structure of defects. 
Here we address this fundamental question by adopting a theoretical 
and experimental approach combining few nanometer resolved cathodoluminescence 
 techniques \cite{Zagonel2010} with high resolution transmission electron 
microscopy images and state of the art quantum mechanical simulations. We show 
how additional excitonic emissions are associated with changes in the layer 
stacking order and how these structures can appear at local layers folds. 

  \begin{figure*}[tb]
  \includegraphics[width=\textwidth]{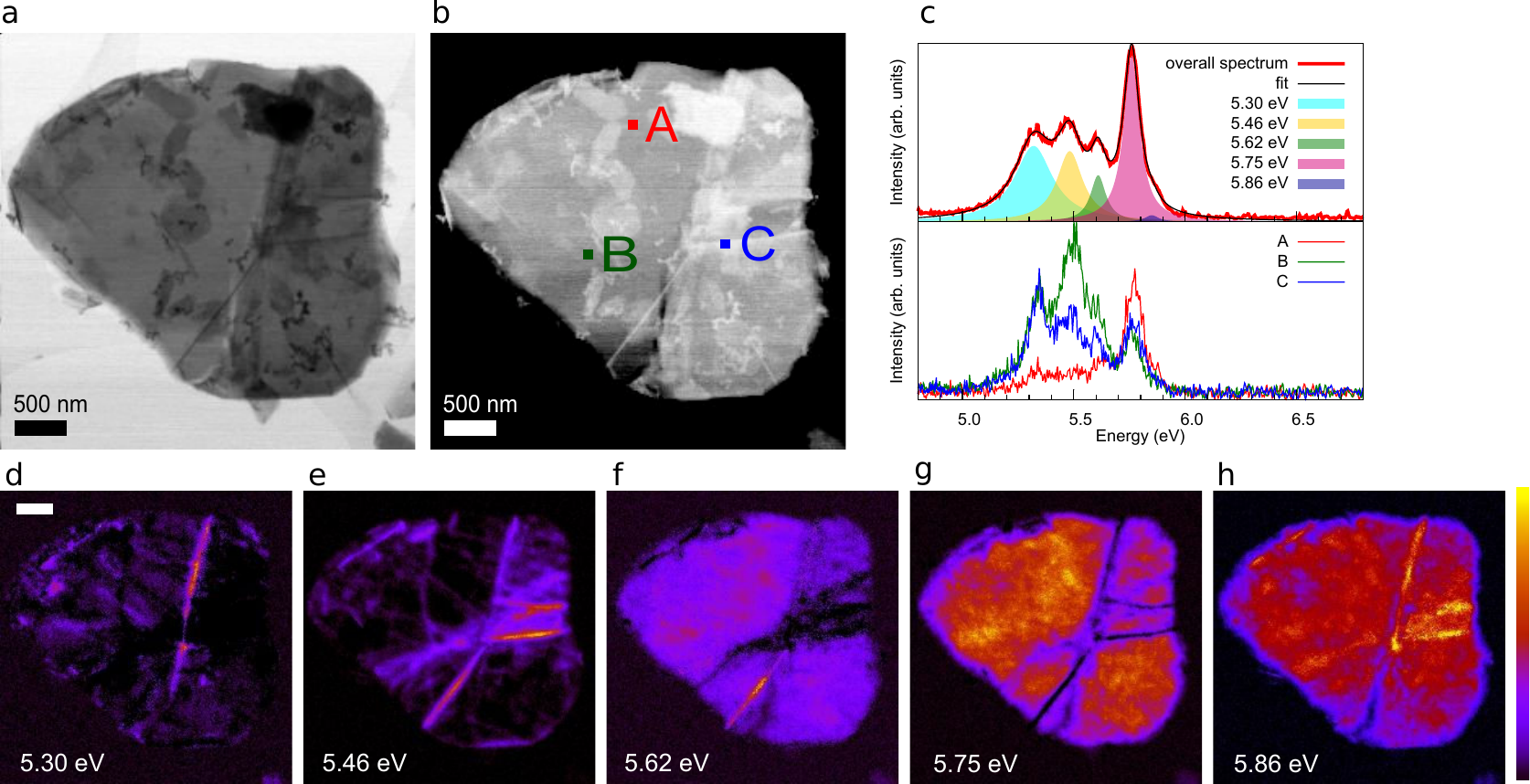}
  \caption{\textbf{a} Bright field and \textbf{b} dark field images of an 
  individual BN flake. \textbf{c} Overall emission spectrum of the flake and 
  individual spectra taken at specific probe positions indicated in panel b. 
  \textbf{d-h} Emission maps for individual emission peak. Intensity is 
  normalized independently within each individual map.}
  \label{emission spectra}
  \end{figure*}

h-BN flakes with an average lateral size of about 2-3 $\mu$m and 
thickness from a few nanometers down to the monolayer were obtained by 
chemical exfoliation through ultrasonication in isopropanol of commercially available 
micrometric powder. Luminescence within 
individual h-BN nanoflakes was investigated at a nanometric resolution 
through  cathodoluminescence hyperspectral imaging. Individual particles were 
scanned by a 1 nm electron probe in a scanning transmission electron 
microscope (STEM) and a full emission spectrum in the 3.0-6.0 eV energy range  
was acquired at each probe position together with a bright field and a high 
angle annular dark field image, HAADF (a full emission spectrum is shown in 
supplementary figure 1). In this work we will focus on excitonic emissions in 
the  5.3-5.9 eV energy range, the most interesting spectral region for optical 
applications \cite{Watanabe2009b}, leaving to future discussion additional 
emissions appearing occasionally in the middle of the band gap.

In figure \ref{emission spectra}.a,b we present bright field and HAADF images 
synchronously acquired with a hyperspectral image ($9\cdot 10^4$ sequential 
spectra obtained while scanning the sample). A series of five 
emission lines at 5.30, 5.46, 5.62, 5.75 and 5.86 eV is clearly visible in the 
overall cathodoluminescence spectrum of the nanoparticle (Fig. \ref{emission 
spectra}.c). These energies correspond to previously reported values for h-BN 
crystals and multi-walled BN nanotubes \cite{Watanabe2009,Jaffrennou2008a}. 
Photoluminescence experiments conducted at different temperatures show finer 
structures for temperatures below 50 K \cite{Watanabe2009}. These additional 
features are smeared  at higher temperature and thus are not visible through our 
experimental set up operating at about 150 K. 

The three example spectra presented in Fig. \ref{emission spectra}.c show the 
strong inhomogeneities in the peak's relative intensities occuring at different 
probe positions. Thin samples present regions in which only the h-BN bulk 
exciton, usually called the free exciton, occurs (supplementary figure 2). In 
order 
to extract the spectral weight of individual emission lines, a multi-Lorentzian 
fitting routine has been applied to each spectrum of the spectrum image. Whereas 
peak energies were free parameters of the fitting procedure, no relevant 
energy shifts were detected  across individual and between different flakes. 
Intensity maps derived from this analysis are shown in figure \ref{emission 
spectra}.d-h. Each map has been normalized independently on the basis of the 
most intense pixel. It should be mentioned however that peaks at 5.62 and 5.86 
eV are systematically significantly weaker than other emission lines and this 
behavior can not be attributed solely to detection efficiency differences.

A detailed analysis of intensity maps demonstrates strong inhomogeneities 
occurring at lines crossing the flakes. The free excitonic emission at 5.75 eV 
and the two emission peaks at 5.62 eV and 5.86 eV are distributed all across the 
particle. Contrarily, the two emission peaks at 5.30 eV and 5.46 eV are mostly 
localized at the lines. All emission maps are clearly not correlated one 
with another and thus all five emission features should be considered 
independent.

Lines visible in CL maps appear also clearly in HAADF and bright field images 
and thus they are associated with local structural changes. Furthermore they 
occur 
mostly in pairs separated by few up to hundred of nanometers (supplementary 
figure 3). For a better understanding on how specific emission peaks 
are associated with specific structural features, these lines have been 
characterized by bright field imaging in an aberration corrected 
STEM. 

\begin{figure*}[tb] 
\includegraphics[width=0.75\textwidth]{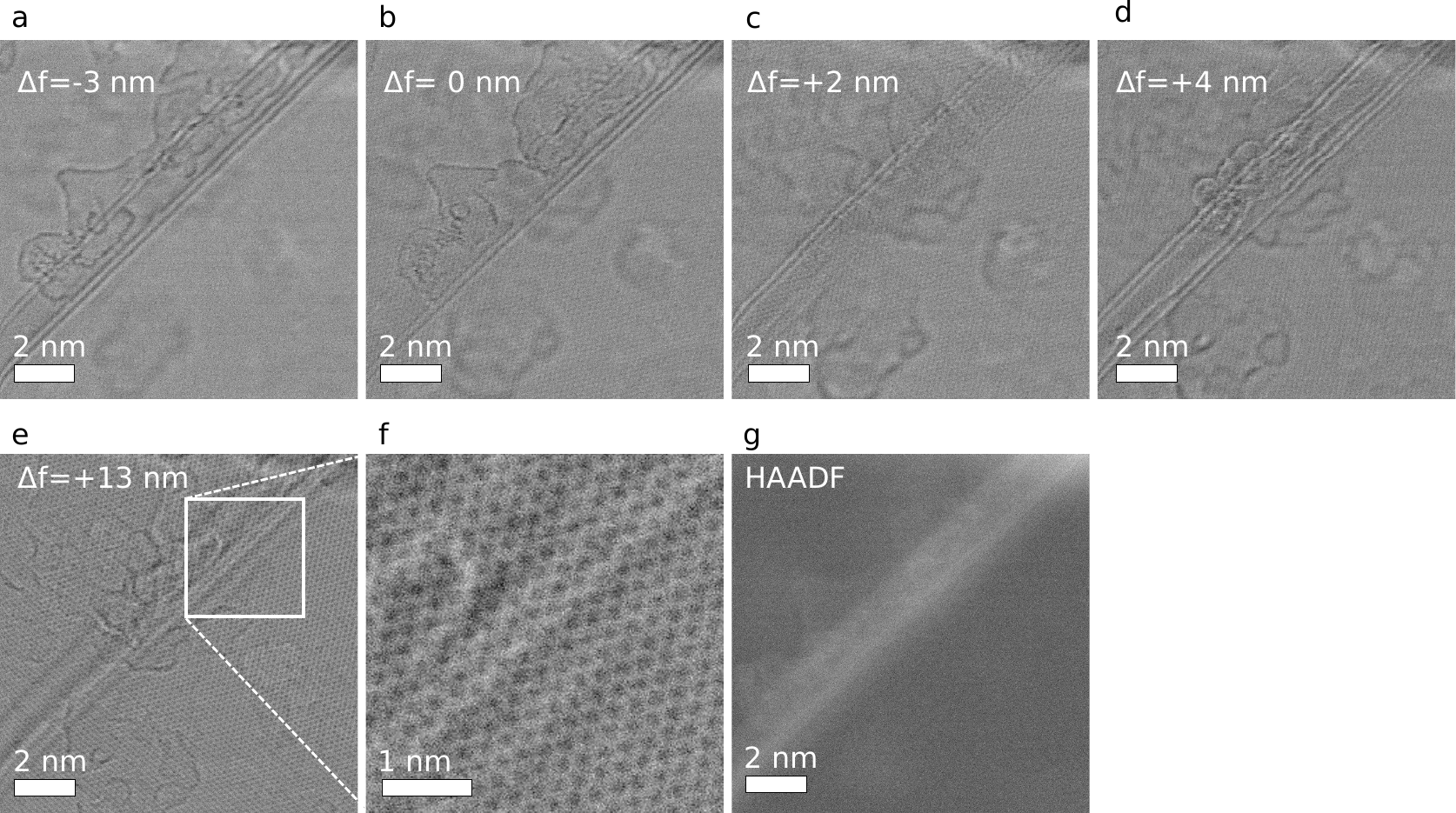}
\caption{\textbf{a-e.} Five micrographs of two couple of parallel fringes 
extracted from a 34-image focal series. The two pair of fringes are 
focussed at different focal depths \textbf{f} Magnified image 
showing that the fringes are aligned along an arm-chair direction of the h-BN 
lattice. \textbf{g.} Dark field image of the region showing a local increase 
of the projected density at the fold.}
\label{focal-series}
\end{figure*}

In figure \ref{focal-series}.g we present a HAADF image for a pair of parallel 
bright lines and the corresponding scanning transmission bright field focal 
series (Fig.\ref{focal-series}.a-f). Each line is formed by two parallel fringes 
indicating layers parallel to the electron beam direction. The under-focused 
image (Fig. 
\ref{focal-series}.a) is rather similar to the image of a double walled BN nanotube, but unlike 
the case of a nanotube the focus for the two 
series of fringes occurs at two different defocus values (Fig. 
\ref{focal-series}.b,c). This corresponds to a situation where the two series of 
fringes lie at different depths. Finally when both pairs of fringes are 
overfocused the image of the underlying h-BN lattice is recovered (Fig. 
\ref{focal-series}.e,f). A similar behavior is found also for structures with a 
higher number of fringes (supplementary figure 4).

\begin{figure*}[tb]
\includegraphics[width=\textwidth]{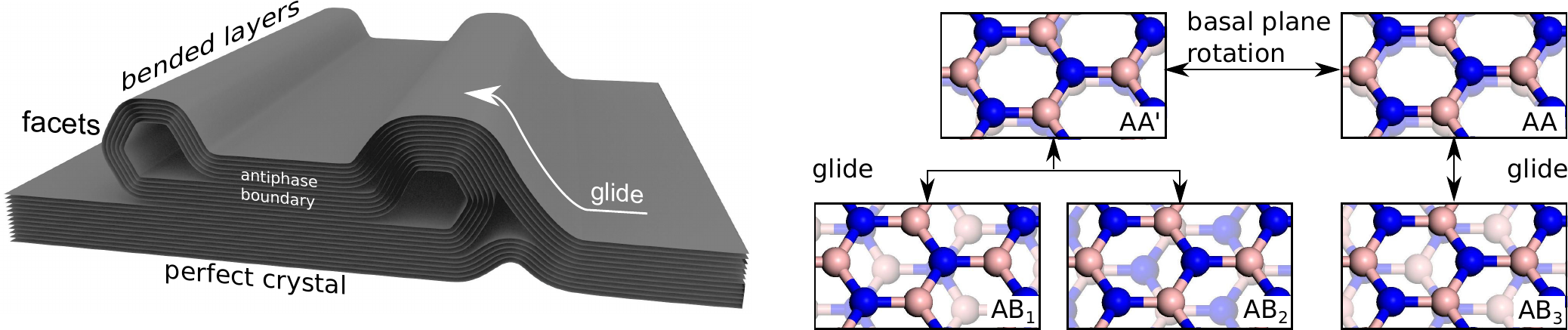}
\caption{\textbf{Left}. Model of a faceted double fold in a h-BN crystal. At different sections of 
the crystal layers glides and antiphase boundaries appear. \textbf{Right}. 
Morphological deformations promote transition to different stacking sequences. 
Layers glide at bended regions transforms the original AA$'$ stacking in the 
AB$_1$ 
or AB$_2$ stackings. AA and AB$_3$ are associated to antiphase boundaries in 
overlapping regions.}
\label{fold}
\end{figure*}

The behavior described here can be explained by a model in which a layer stack 
can deform forming two almost parallel folds (see Fig. \ref{fold}). This double 
folded structure has been proposed to occur in neutron irradiated graphite 
\cite{Heggie2011} and it has been recently imaged in CVD grown few-layer graphene 
\cite{Robertson2011,Kim2011} and MoS$_2$ flakes \cite{Castellanos-Gomez-13}. 
Our observations do not allow to exclude the additional presence of single folds whereas they 
might be less common than double folds. Emission peaks at energies of 5.30 eV and 5.46 eV are thus 
solely associated with folds.

The different curvature of the layers at the fold induces, as for any multi 
layered structure, a relative glide of the layers and thus a loss of the 
original AA$'$ stacking sequence of the perfect h-BN crystal (Fig. \ref{fold}). 
The local atomic structure of the folds is thus analogous to the structure of 
multi walled nanotubes where in principle no defined stacking sequence exists. 
However, in multiwalled BN nanotubes and in other inorganic nanotubes the competition between 
bending and 
interplane energies tends to localize the curvature and to promote polygonal 
tube sections with facets of well defined stacking, not necessarly AA$'$ 
\cite{Celik2005,Tibbetts2012,Golberg2007a}. The local atomic structure at the fold's facets
should thus correspond to stacking sequences that can be generated from the original 
AA$'$ structure by a glide of the layers. Two metastable configurations can then 
be obtained by translating one of the basal planes by a vector $t\cdot(1/3,2/3,0)$, with $t$ a real 
number.
For $t=1$ the $\text{AB}_1$ structure is obtained, where, following an analogous nomenclature for 
graphite nitrogen and boron atoms lies at $\alpha$ and $\beta$ sites 
respectively, and for $t=2$, the  $\text{AB}_2$ structure, where atoms positions are inverted 
(Fig. \ref{fold}) \cite{Maron10,Yin2011,Constantinescu13}. If the fold follows an 
armchair direction, as is seen most often in scanning transmission bright field 
images (Fig. \ref{focal-series}.f,g), antiphase domains appear in the region 
between the folds. The stacking of planes at antiphase domain boundaries 
corresponds to a metastable AA order or, by applying an additional translation, 
an $\text{AB}_3$ order (Fig. \ref{fold}). Besides the specific fold case, 
antiphase domains are, as for graphite, a common defect in h-BN and indeed AA 
stacking orders have been recently identified in few layer h-BN crystals 
\cite{Shmeliov2013}.

These topological considerations suggest that additional excitonic peaks should 
be associated solely with changes of the crystal symmetry in the above-mentioned 
five possible stacking sequences. In order to confirm this hypothesis we have 
performed a complementary theoretical investigation of the optical spectra for 
these configurations. It has been successfully demonstrated that fundamental 
insights on emission can be provided by an analysis of simulated absorption 
spectra, taking into account that absolute energies could not be accessed due to 
the neglect of Stokes shifts \cite{Wirtz2005, Wirtz2008,Watanabe2009}. Wave 
functions were obtained in the framework of density functional theory using the 
local density approximation and quasi particle energies were corrected by a 
subsequent GW treatment \cite{Hedin65}. In a first step, optical spectra were evaluated in 
the random phase approximation (RPA). Finally, electron hole interactions, 
associated with excitonic effects that dominate the h-BN optical spectra, have 
been included by solving the Bethe-Salpeter equation (BSE) \cite{Onida02}. This 
computational scheme has been successfully applied to the study of the 
absorption spectrum of h-BN and BN nanotubes \cite{Wirtz2005, Wirtz2008, 
Attaccalite2011, Galambosi2011a}. Due to the well known underestimation of the 
h-BN band gap in the non self consistent GW computational scheme a 0.35 eV rigid 
shift has been applied to all spectra that allows the alignment of the excitonic 
peak of the AA$'$ stacking to the known experimental value. 

\begin{figure}[b] 
\includegraphics[width=0.70\columnwidth]{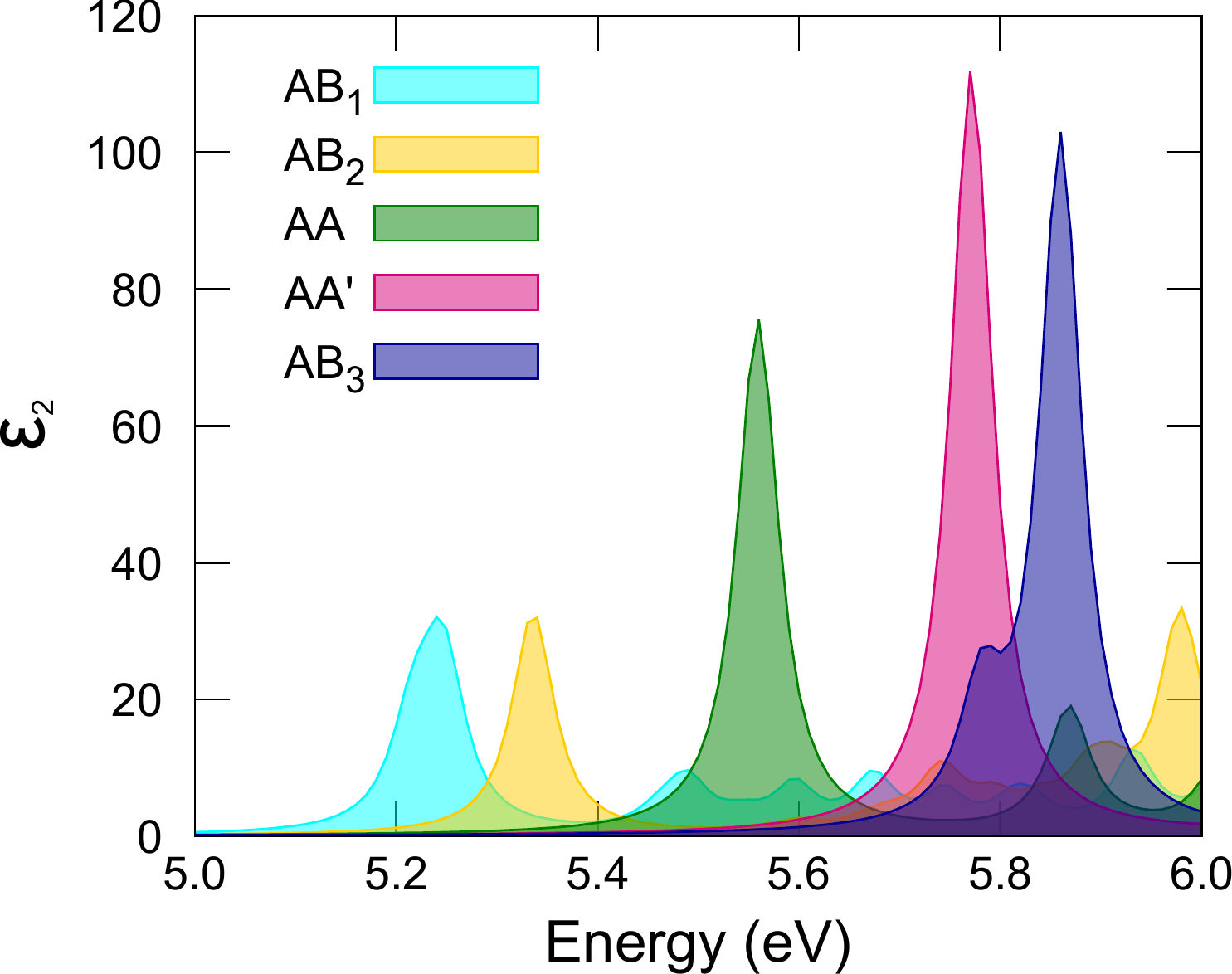}
\caption{Imaginary part of the dielectric function (optical absorption spectum) 
 for different h-BN stacking obtained in the framework of GW+BSE calculations. }
\label{theo}
\end{figure}

For different h-BN polytypes, bands dispersion remains invariant in the \textbf{$a^*$} 
\textbf{$b^*$} plane. However for directions parallel to the 
 \textbf{$c^*$} axis the last two occupied and first two unoccupied bands are either degenerate or 
split depending on 
stacking order. This leads to changes in the direct band gap 
observed both at the DFT and GW level (see Table 1 in the supplementary materials). 
This behavior has also a fundamental role in defining the optical properties of different 
stacking configurations. Indeed in agreement with the 
work of Arnaud et al. on AA$'$ h-BN \cite{Arnaud2006a}, we find for all 
configurations that directions parallel to the 
\textbf{$c^*$} axis give rise to the first absorption peak of $\varepsilon_2$  
both at RPA and BSE levels. Quasi-particle band gaps and exciton binding 
energies are finally strongly dependent on the stacking order and determine the 
energies of optically active excitons (more details will be provided in a future work). 

All absorption spectra presented in Fig. \ref{theo} show strong excitons with 
well separated energies in the range between 5.2 to 5.9 eV depending on the 
stacking configuration. Taking as a reference the standard  AA$'$ exciton 
at 5.75 eV, three additional excitons, associated with the $\text{AB}_1$, 
$\text{AB}_2$ and AA configurations, occur at lower energies whereas that of the 
$\text{AB}_3$ configuration is found at a higher one. Energy range and peak 
distribution coincide with experimental results. Thus theoretical simulations 
enable us to associate each of the five measured emission peaks with the five 
high symmetry configurations analyzed. In  particular emissions at 5.30 and 5.46 
eV, strongly localized at folds, can be attributed to $\text{AB}_1$ and 
$\text{AB}_2$ stacking sequences respectively. These models are compatible with 
the faceted fold structure where discret layer glides between facets give 
rise to well defined stacking orders. Emissions at 5.62 and 5.86 eV 
can be associated with the AA and $\text{AB}_3$ stacking respectively. Since 
these stackings appear at antiphase boundaries, these excitons occur in large 
regions of the flake and their signal is weak due to the small number of atomic 
planes involved. We underline that each intermediate stacking gives rise to 
different exciton energies (supplementary figure 5). The experimental evidence 
for only five discrete emission lines is thus a clear indication of a finite 
number of stacking configurations and it confirms the hypothesis of faceted 
folds.

In conclusion, using a complementary experimental and theoretical approach we 
have provided a microscopic explanation for the complex emission spectrum of 
h-BN relating local changes in the layer stacking order to the appearance of 
additional excitons. Furthermore, we have shown how these defects and the 
associated excitonic emissions can be strongly localized at folds crossing the 
flakes. Finally our theoretical analysis reveals that, whereas additional 
emission lines in h-BN can still be interpreted as excitons bound to defects, 
they can  be better described as bulk excitons of different polytypes. Recently, 
h-BN polytypes have been discussed from a morphological and energetic point of 
view but little attention was paid to their optical and electronic properties 
\cite{Maron10,Yin2011,Constantinescu13,Shmeliov2013}. This study emphasizes the 
role of stacking changes in defining the optical properties of h-BN. The 
importance of stacking faults is thus directly related to the interest in h-BN 
as one of the most promising materials for high performance far ultraviolet 
emitters.

\acknowledgments

The authors acknowledge support from the Agence Nationale de la Recherche 
(ANR), program of future investment TEMPOS- CHROMATEM (No. 
ANR-10-EQPX- 50), and Triangle de la Physique, Theo-STEM 
project (No. 2010-085T). The work has also received funding from the 
European Union in Seventh Framework Programme (No. FP7/2007- 2013) under Grant 
Agreement No. n312483 (ESTEEM2) and Marie Curie Intra-European Fellowship No. 
326794 (EXPRESS).

\part{Supplementary informations}

 \setcounter{figure}{0}

\section*{Synthesis}

Few-layer hexagonal boron nitride flakes was obtained by chemical exfoliation 
following the protocol presented in Ref \cite{Coleman2011a}. A solution of 10 
mg 
of h-BN powder in 10 ml of isopropanol was sonicated for 7 hours and 
subsequently centrifuged for 120 minutes at 500 revolutions per minute. The 
supernatant was then dropped onto a TEM copper grid. Samples were then purified 
by a three hour 500 $^\circ$C thermal treatment in a 800 mbar forming gas 
atmosphere (95\% N$_2$, 5\% H$_2$). HREM images and core level electron energy 
loss spectroscopy confirmed the absence of any carbon contamination due to 
solvent residuals. Flakes presented a lateral size of few microns.

\section*{Microscopy and spectroscopy}

Cathodoluminescence was performed in a dedicated VG-HB501 scanning 
transmission electron microscope operating at 60 keV, below the atomic 
displacement threshold for h-BN  in order to avoid irradiation 
damage \cite{Zobelli2007b,Kotakoski2010}.
Optical spectra were collected using an optical spectrometer with a 300 
groove diffraction grating blazed at 300 nm. The resolution of the CCD was 0.17 
nm/pixel. 

High resolution scanning transmission electron microscopy images were
recorded in a NION ULTRA-STEM 200 microscope operating at 60 keV. Lattice 
planes were visible in bright field images but not in dark field due to the 
resolution loss at low voltage and the rather high thickness and orientation 
complexity of the sample.

\subsection*{Cathodoluminescence}

\begin{figure}[tb]
\includegraphics[width=0.8\columnwidth]{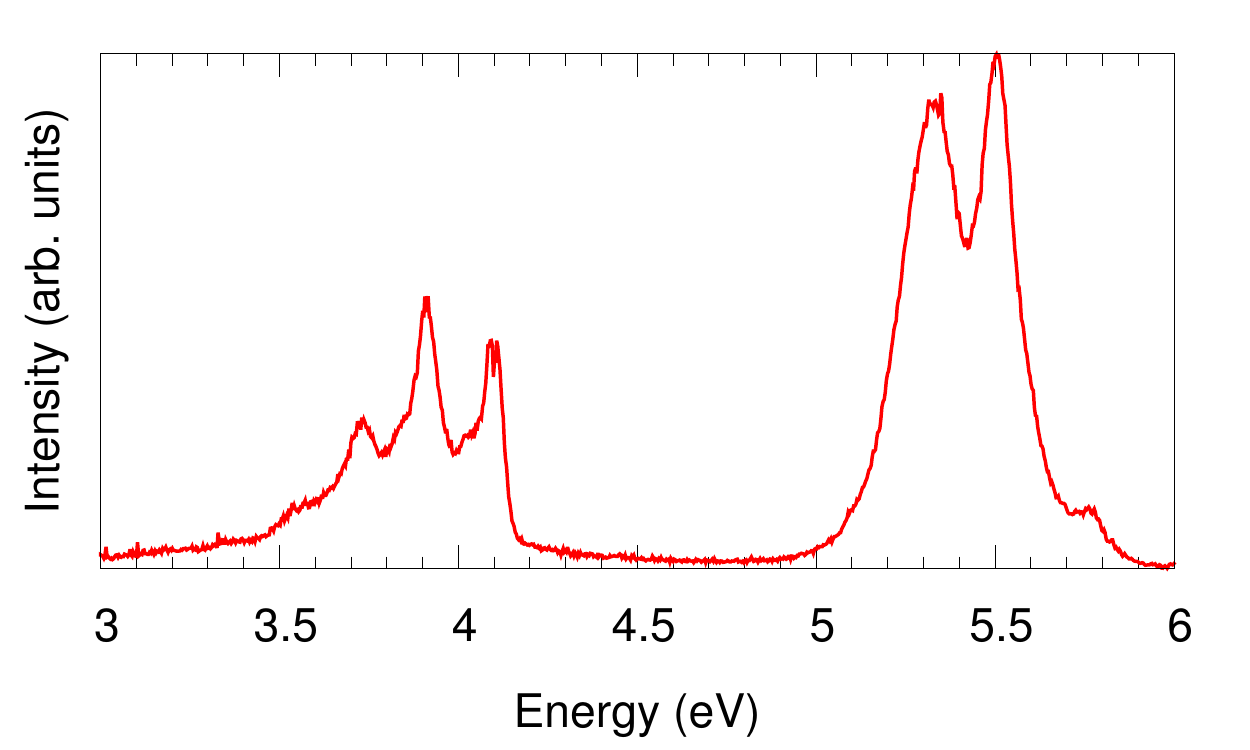}
\caption{Full emission spectrum recorded for an h-BN flake on which 
additional intraband emission appears.}
\label{Full spectra}
\end{figure}

In figure \ref{Full spectra} is shown a full emission spectrum of a h-BN 
flake in the energy range from 3 to 6 eV. Besides high energy emission peaks 
commented in the main article, an additional intraband broad emission appears in 
the energy range 3.2-4.5 eV on which three sharp peaks at 3.73, 3.90 and 4.09 eV
are superposed. These additional emission features occur occasionally in 
limited regions of the flakes. Previous photoluminescence studies have 
attributed the broad emission to donor acceptor pairs and the sharp lines to 
phonon replicas of a luminescent impurity \cite{Museur2008}. The spatial 
localization of these signals obtained through cathodominescence will be 
discussed in a future work.

\begin{figure}[tb]
\includegraphics[width=0.8\columnwidth]{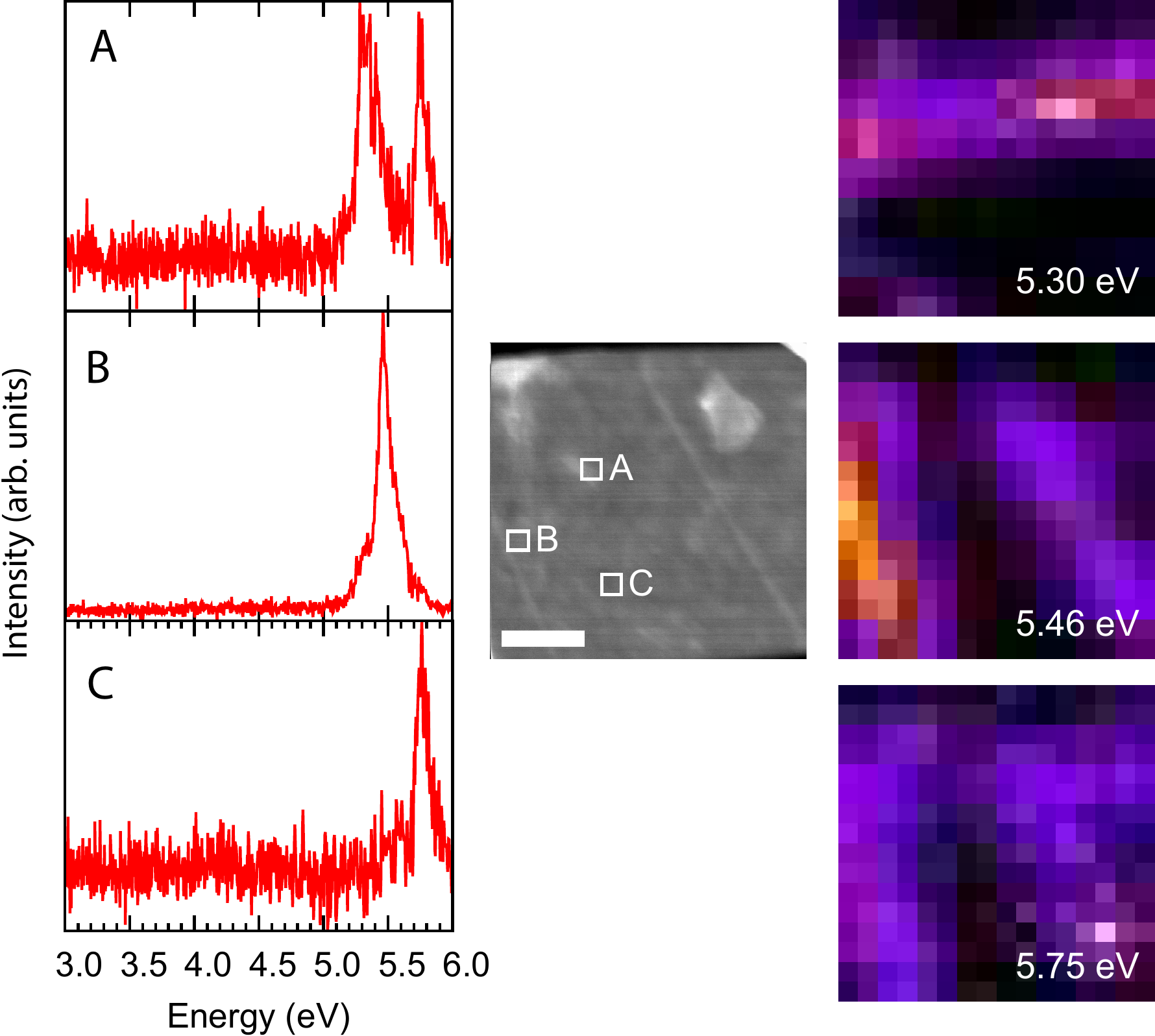}
\caption{\textbf{Left.} Spectra extrated from a cathodoluminescence 
hyperspectral image for probe position indicated in the   
\textbf{middle} HAADF image . \textbf{Right.} Cathodoluminescence emission maps 
of the 5.30, 5.46 and 5.75 eV emission peaks.}
\label{thin-region}
\end{figure}

In figure \ref{thin-region} we present cathodoluminescence maps obtained for a 
thin h-BN flake. Spectra extracted from the original hyperspectral image show 
that single emission lines can appear in different regions of the flake. In the
spectrum C it is solely visible the 5.75 eV emission line characteristic of a 
perfect h-BN crystal. This indicates that thin regions can be defect free and 
have the standard AA$'$ stacking of h-BN.

\subsection*{Microscopy images}

\begin{figure*}[h]
\includegraphics[width=\textwidth]{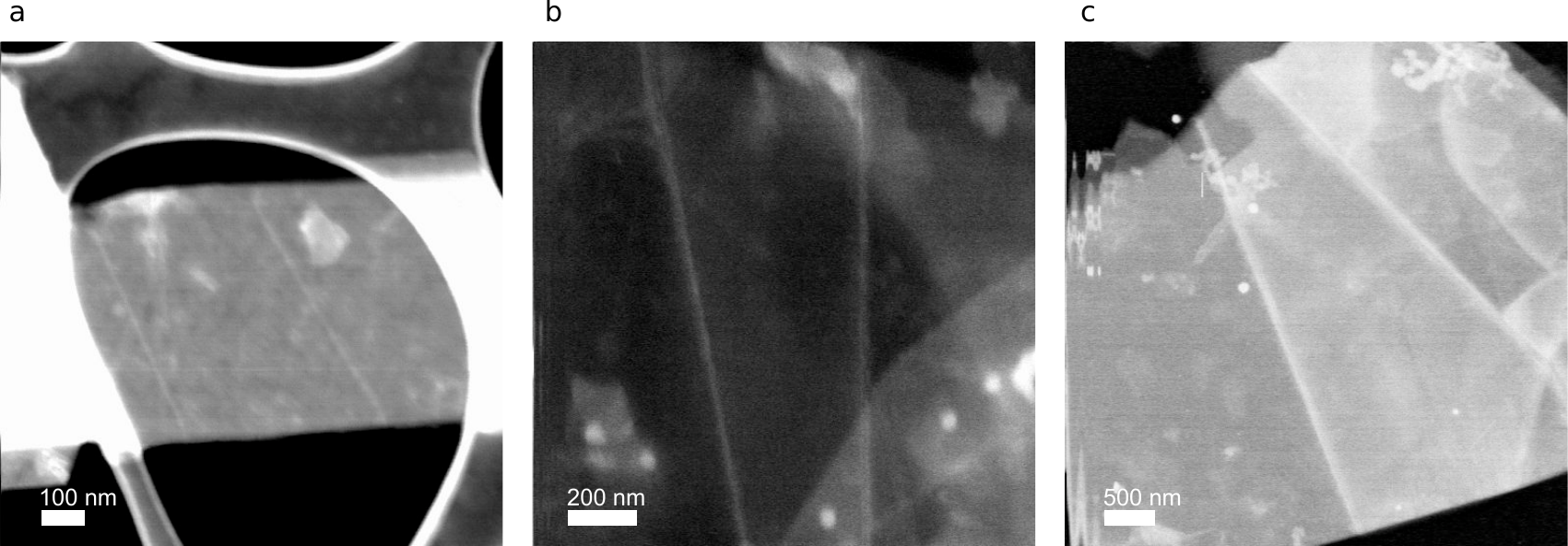}
\caption{Low magnification HAADF images for different h-BN flakes where a 
series of almost parallel bright lines appear.}
\label{lowmag-HAADF}
\end{figure*}

\begin{figure*}[htb]
\includegraphics[width=\textwidth]{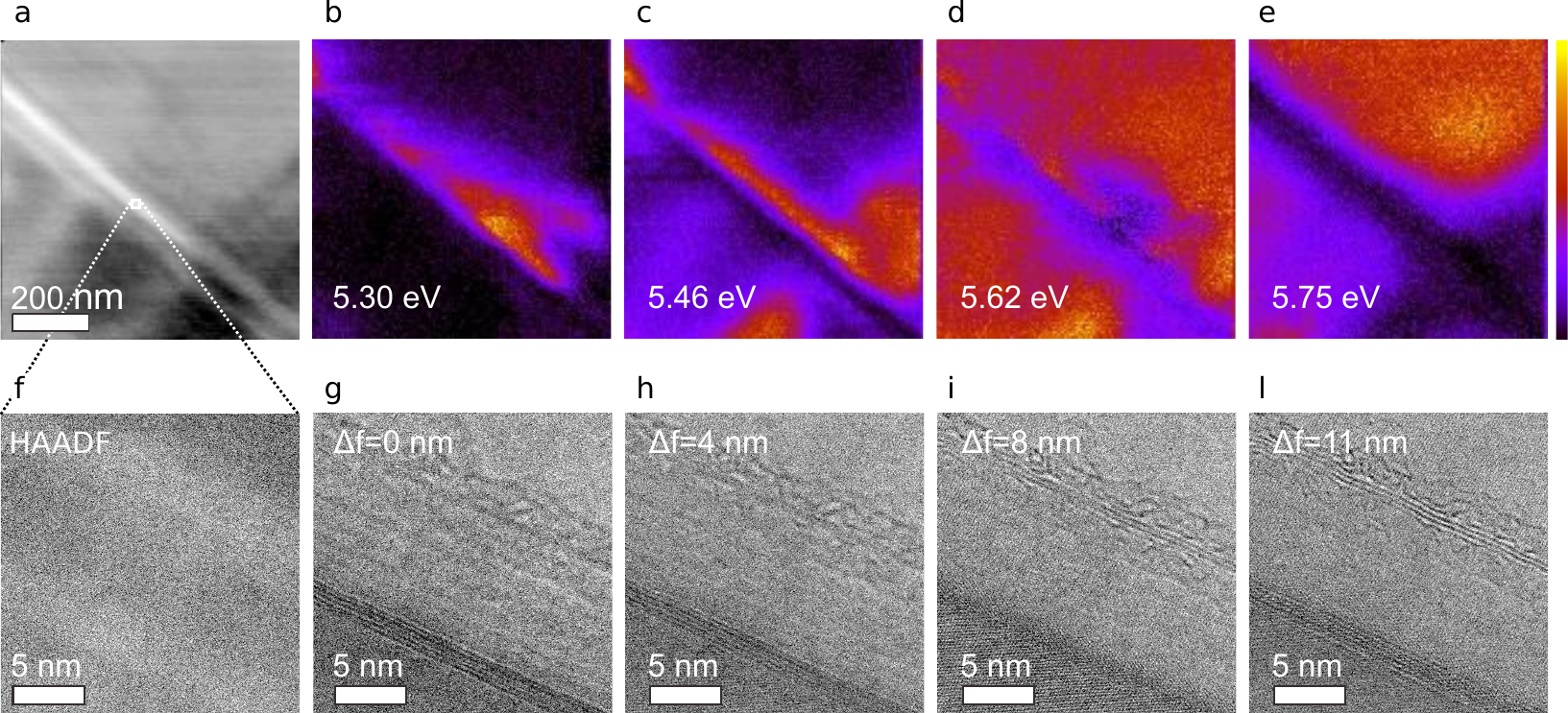}
\caption{\textbf{a.} Low magnification HAADF image of a line in a h-BN flake 
and \textbf{b-e} cathodoluminescence maps of the region corresponding to 
different excitons. \textbf{f.} Higher magnification HAADF 
image of the same zone and four bright field micrographs extracted from a 40 
image focal series.}
\label{emission spectra}
\end{figure*}

Figure \ref{lowmag-HAADF} shows low magnification scanning transmission 
electron microscopy (STEM) high angle annular dark field images  (HAADF) of 
three h-BN flakes where two bright lines separated by few hundred nanometers 
are visible. The structure is similar to previous observations of double folds 
in few-layer CVD grown graphene \cite{Robertson2011,Kim2011}. As discussed by 
Robertson et al. \cite{Robertson2011} folds can be parallel (as in figure  
\ref{lowmag-HAADF}.a) or form a narrow angle depending on the fold's directions 
with respect to the graphene lattice. In figure \ref{lowmag-HAADF}.c a local 
increase of the projected density in the region between the two lines is 
visible. This local thickening of the flake is compatible with the double fold 
model proposed in this work (see figure 3 of the article).

In figure \ref{emission spectra}  we present (Fig. \ref{emission spectra}.a) a 
low magnification HAADF image of a bright line and (Fig. \ref{emission 
spectra}.b-e) emission maps of four excitonic peaks extracted from a 
hyperspectral cathodoluminescence image. As discussed in the article, peaks at 
5.30 and 5.46 eV are strongly localized at the folds. At a higher 
magnification the bright line appears as being formed by two parallel lines 
separated by about 10 nm (Fig. \ref{emission spectra}.f). The same region has 
been imaged through a high resolution bright field focal series obtained in an 
aberration corrected scanning transmission electron microscope (Fig. 
\ref{emission spectra}.g-l). Each line of the dark field image is constituted of 
five parallel fringes indicating layers parallel to the electron beam direction. 
The focus for the two series of fringes occurs at two different defocus values. 
The structure can thus be interpreted as a double fold involving five h-BN 
planes.

\section*{Theory}

Quantum mechanical calculations of electronic and optical properties for the 
five stacking configurations of bulk h-BN were performed following a well 
established computational procedure. Ground state electronic structures were 
derived using plane waves pseudo-potential density functional theory as 
implemented in the Abinit package within the local density approximation (LDA). 
A Monkhorst-pack grid of $10 \times 10 \times 4$ $k$-point was used to sample 
the Brillouin zone of the system. An energy cutoff of 32 Ha was demonstrated to 
be enough for the convergence of the total energy. Following the ground state 
calculations, self energy corrections to the Kohn-Sham eigenvalues for high 
symmetry points of the Brillouin zone was obtained using many body perturbation 
theory (MBPT) adopting a non self-consistent GW approach. The $6 \times 6 \times 
2$  $k$-grid for GW calculations was not shifted meaning that the Gamma point 
was included. Since for all the structures the dispersion of the GW correction was
lower than 0.5 eV, a scissor operator, chosen at the Gamma point, has been
used for the optical spectra calculation.

In a first step, optical spectra were evaluated in the random phase 
approximation (RPA). Dielectric functions were derived using a $15 \times 15 
\times 7$ shifted k-grid and 30 energy bands. Crystal local field effects were 
included (with 50 G vectors) and a scissor operator was applied to 
take into account GW self energy corrections. In a second step, we performed optical spectra 
calculations by using the Bethe-Salpeter equation, which permits to include electron-hole 
interaction by mixing all the electronic transitions in the resolution of a two 
particle equation. A $15 \times 15 \times 7$  shifted $k$-grid was used, with 
50 G-vectors and a broadening of 0.025 eV.

\begin{table*}[htb]
 \setlength{\tabcolsep}{12pt}
 \begin{tabular}{c|c|c|c|c|c}
 \hline
 \hline
Configuration & LDA   & GW correction& RPA  & First optical active 
& 
Exciton binding \\
 & direct gap & at the $\Gamma$ point & optical gap &  exciton transition energy & energy \\
 \hline
AB$_1$   & 3.51 & 1.72 & 5.23 & 4.89 & 0.34 \\
AB$_2$   & 3.70 & 1.75 & 5.45 & 4.98 & 0.47 \\
AA       & 2.95 & 1.72 & 5.70 & 5.21 & 0.49 \\
AA$'$      & 4.52 & 1.67 & 6.19 & 5.42 & 0.77 \\
AB$_3$   & 4.35 & 1.74 & 6.09 & 5.43 & 0.66 \\
 \hline
 \hline
 \end{tabular}
 \caption{DFT-LDA direct gap, GW correction at the $\Gamma$ point, RPA optical 
gap, BSE first optical active exciton energy and 
exciton binding energy calculated for different h-BN stacking configurations. 
All energies are expressed in eV.}
 \label{table-gaps}
\end{table*}

\begin{figure*}[htb]
\includegraphics[width=0.8\textwidth]{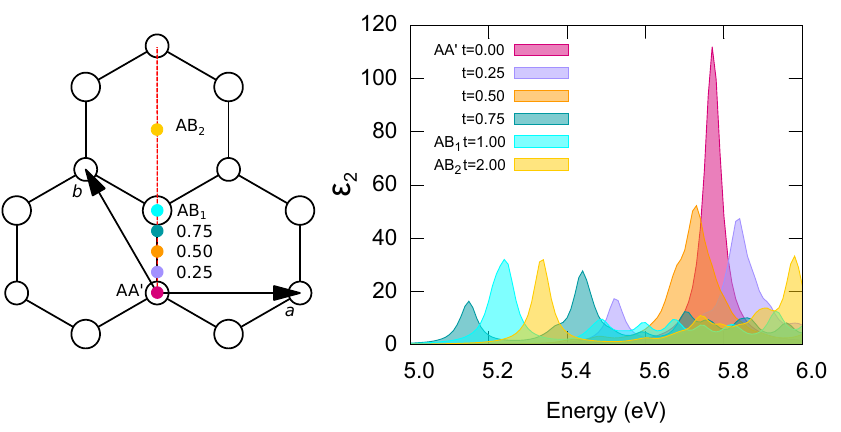}
\caption{\textbf{Left} Schematic representation of the transformation of 
AA$'$ h-BN in the AB$_1$, AB$_2$ and intermediate stacking configurations
obtained by gliding one basal plane  by a vector $t\cdot(1/3,2/3,0)$.
\textbf{Right} Imaginary part of the dielectric function (optical absorption 
spectum) for different h-BN stacking derived in the framework of GW+BSE 
calculations.
}
\label{translations}
\end{figure*}

In table \ref{table-gaps} we report LDA direct band
gaps, quasi particle corrections at the gamma point (scissor operator applied), optical gap and 
energy of the first optical active exciton for the five h-BN polytypes 
considered. Both band gaps and exciton binding energies contribute to 
the determination of the energy of the first optical active exciton.

The two metastable configurations AB$_1$ and AB$_2$ can be obtained from the 
$\text{AA}'$ structure by translating one of the basal planes by a vector 
$t\cdot(1/3,2/3,0)$ where $t$ assumes the value 1 and 2 respectively (Fig. 
\ref{translations}). In order to investigate the effect on the optical 
properties of a continuum variation between high symmetry configurations to model the glide, 
absorption spectra have been computed following a GW+BSE scheme for three 
additional configurations with t=0.25, 0.5 and 0.75. In Fig. \ref{translations} 
it is shown that a different absorption spectrum is associated with each 
configuration. The experimental observation of only five excitonic emission 
peaks is thus a clear indication for a limited number of high symmetry stacking 
configurations present in the crystals.

\end{document}